\documentclass[twocolumn,aps,superscriptaddress,prc,reprint]{revtex4-1}
\usepackage{blindtext}
\usepackage{graphicx}
\usepackage{dcolumn}
\usepackage{bm}
\usepackage{booktabs}
\usepackage{color}
\usepackage{mathrsfs}
\usepackage{ulem}
\usepackage{epstopdf}
\usepackage{verbatim}
\usepackage{indentfirst}
\usepackage{float}
\usepackage[colorlinks,
           bookmarks=true,
           linkcolor=blue,
            =blue,
            anchorcolor=black,
            citecolor=blue
            ]{hyperref}
\usepackage[bf,raggedright,indentafter]{titlesec}
\usepackage{hypcap}
\usepackage{enumerate}
\usepackage{subfigure}

\begin{document}
\title{Baryon-strangeness Correlations in Au+Au Collisions at $\sqrt{s_\mathrm{NN}}$=7.7-200 GeV from the UrQMD model}

\author{Zhenzhen Yang}
\affiliation{Key Laboratory of Quark and Lepton Physics (MOE) and Institute of Particle Physics,\\
Central China Normal University, Wuhan 430079, China}
\author{Xiaofeng Luo}
\email[Electronic address:~]{xfluo@mail.ccnu.edu.cn}
\affiliation{Key Laboratory of Quark and Lepton Physics (MOE) and Institute of Particle Physics,\\
Central China Normal University, Wuhan 430079, China}
\affiliation{Department of Physics and Astronomy, University of California, Los Angeles, California 90095, USA}
\author{Bedangadas Mohanty}
\affiliation{School of Physical Sciences, National Institute of Science Education and Research, HBNI, Jatni 752050, India }
\begin{abstract}
Fluctuations and correlations of conserved charges are sensitive observables for studying the QCD phase transition and critical point in high-energy heavy-ion collisions. We have studied the centrality and energy dependence of mixed-cumulants (up to fourth order) between net-baryon and net-strangeness in Au+Au collisions at $\sqrt{s_{NN}}$= 7.7, 11.5, 19.6, 27, 39, 62.4, 200 GeV from UrQMD model.  To compare with other theoretical calculations, we normalize these mixed-cumulants  by various order cumulants of net-strangeness distributions. We found that the results obtained from UrQMD calculations are comparable with the results from Lattice QCD at low temperature and hadron resonance gas model. The ratios of mixed-cumulants  ($R_{11}^{BS},R_{13}^{BS},R_{22}^{BS}$,~$R_{31}^{BS}$) from UrQMD calculations show weak centrality dependence. However, the mixed-cumulant ratios $R_{11}^{BS}$ and $R_{31}^{BS}$ show strong increase at low energy, while the $R_{13}^{BS}$ snd $R_{22}^{BS}$ are similar at different energies.  Furthermore, we have also studied the correlations between different hadron species and their contributions to the net-baryon and net-strangeness correlations. These models studies can provide baselines for searching for the signals of QCD phase transition and critical point in heavy-ion collisions.

\end{abstract}
\maketitle

\section{Introduction}
Fluctuations and correlations of the conserved charges have been proposed to be sensitive observables to study the QCD phase transition and Critical Point (CP) in relativistic heavy-ion collisions~\cite{Misha,Volker, Asakawa,science,flu_review}. Theoretical calculations show that fluctuations and correlations  of conserved charges are distinctly different in the hadronic and Quark Gluon Plasma (QGP) phases~\cite{1511.06541}. In the deconfined state of quarks and gluon,  the elementary set of conserved charges are given by the corresponding quark flavors: net upness($\Delta{u}$=u-$\overline{u}$), net downess($\Delta{d}$=d-$\overline{d}$)and net strangeness($\Delta{s}$=s-$\overline{s}$)~\cite{1309.2317,014004,PRL109,JHEP1201}. In hadronic state the conserved charges are net-baryon (B), net-charge (Q) and net-strangeness (S). Recently, Lattice QCD calculations have shown that various order (up to fourth order) net-strangeness (S) fluctuations and their correlations with net-charge (Q) and net-baryon (B) are quite sensitive to the quark-hadron phase transition~\cite{PRL109}.

In the years 2010$\sim $2014, Relativistic Heavy Ion Collider (RHIC) has completed the first phase of the beam energy scan (BES) program~\cite{BESI}. In this program the the collision energies in Au+Au collisions were tuned to explore the QCD phase structure at high baryon density~\cite{flu_PRL_STAR,CPOD2014_XFLUO,QM2015_XFLUO}. The STAR experiment at RHIC has collected Au+Au collisions data with $\sqrt{s_{NN}}$ = 7.7, 11.5, 14.5, 19.6, 27, 39, 62.4, 200 GeV, respectively. One of the experimental motivation is to measure the correlations between net-baryon and net-strangeness with these experimental data collected to study the QCD phase transition in heavy-ion collisions~\cite{074043}. In order to provide baselines for the correlation measurements to study the phase transition, we have calculated these observables in Au+Au collisions at RHIC BES energies with UrQMD model and compared them to the model results from Lattice QCD and hadron resonance gas model.  In addition given at lower energies we expect to see the effect of baryon stopping and associated production of strangeness, we have also studied the contributions of various hadron species to the baryon-strangeness correlations.

The paper is organised as follows. In the next section, we briefly review about fluctuations and correlations of conserved charges in hadronic and quark-gluon plasma states and give the corresponding results from hadronic resonance gas model and Lattice QCD. In section III, we present the mixed-cumulant ratios of net-baryon and net-strangeness and also discuss the contribution of strange baryons to the baryon-strangeness correlations in Au+Au collisions at RHIC BES energies from UrQMD model. Then, we give a brief summary in section IV. Finally, we present as an appendix  the statistical error calculations for various order mixed-cumulants.

\section{UrQMD Model}
The UrQMD (Ultra relativistic Quantum Molecular Dynamics) model~\cite{PRC62,PRC73} is based on 
a microscopic transport theory where the phase space description of the
reactions are considered. It treats the propagation of all hadrons on classical 
trajectories in combination with stochastic binary scattering, color string
formation and resonance decay. It incorporates baryon-baryon, meson-baryon and 
meson-meson interactions, the collisional term includes more than 50 baryon species 
and 45 meson species. The model preserves the conservation of electric charge, baryon
number and strangeness number as expected for a QCD matter. It also models the 
phenomena of baryon-stopping an essential feature encountered in high energy heavy-ion
collisions at lower beam energies. The model does not include quark-hadron phase 
transition or the QCD critical point. It can simulate heavy-ion collisions in the energy range from SIS (SchwerIonen Heavy-ion Synchrotron) to Relativistic Heavy Ion Collider (RHIC) and even for the heavy-ion collisions at the Large Hadron Collider(LHC) ~\cite{1999,1998,0805.0567}.
In this model, the space-time evolution of the fireball is studied in terms of excitation  and fragmentation of color strings and formation and decay of hadronic resonances. The comparison of the data (this paper deals with B-S correlations) onto those obtained from UrQMD model will tell about the contribution from the hadronic phase and its associated processes.

\section{Observables}
In statistics, the generating function of mixed-cumulants for random variables, $X_{1}$, $X_{2}$, ..., $X_{n}(n>2)$ are defined as:
\begin{eqnarray}
G(t_{1}, t_{2}, ... ,t_{n})=\log\langle{e^{\sum_{i=1}^n(t_{j}X_{j})}}\rangle.
\end{eqnarray}
Then, the mixed-cumulants of random variables $X_{1}$, $X_{2}$, ..., $X_{n}(n>2)$ can be expressed as~\cite{mixed-cumulants}:
\begin{widetext}
\begin{eqnarray}\label{eq:mixed-moments}
C(X_1,X_2,...,X_n)\!=\!\sum_{\pi}(|\pi|-1)!(-1)^{(|\pi|-1)}\prod_{C\in{\pi}}E(\prod_{i\in{C}}X_i)
=\sum_{\pi}(|\pi|-1)!(-1)^{(|\pi|-1)}\prod_{C\in{\pi},|C|\geq2}E(\prod_{i\in{C}}\delta{X_i}).
\end{eqnarray}
\end{widetext}
where $\pi$ runs through the list of all partitions of {1,2,...n}, C runs through the list of all blocks of partitions $\pi$, $|\pi|$ is the number of parts in the partition and $|C|$ is the number of parts in the block $C$. \\

In mixed-cumulants analysis, we use $B, S$ to represent the net-baryon number and net-strangeness in one event. The average value over whole event ensemble is denoted by $\langle{B}\rangle$, $\langle{S}\rangle$. The deviation of B, S from their mean value are expressed by $\delta{B=B-\langle{B}\rangle}$, $\delta{S=S-\langle{S}\rangle}$, respectively.
Thus, according to Eq. \ref{eq:mixed-moments}, the various order mixed-cumulants of event-by-event distributions of the two random variables B and S are defined as:
\begin{widetext}
\begin{eqnarray}
C(S,S)&=&C_{2}^{S}=\langle{(\delta{S})^2}\rangle,\\
C(B,S)&=&C_{1 1}^{BS}=\langle{\delta{B}\delta{S}}\rangle,\\
C(S,S,S,S)&=&C_{4}^{S}=\langle{(\delta{S})^4}\rangle
-3\langle{(\delta{S})^2}\rangle,\\
C(B,S,S,S)&=&C_{1 3}^{BS}
=\langle{\delta{B}(\delta{S})^3}\rangle
-3\langle{\delta{B}\delta{S}}\rangle\langle{(\delta{S})^2}\rangle,\\
C(B,B,S,S)&=&C_{2 2}^{BS}=\langle{(\delta{B})^2(\delta{S})^2}\rangle
-2\langle(\delta{B}\delta{S})^2\rangle-\langle{(\delta{B}})^2\rangle\langle{(\delta{S}})^2\rangle,\\
C(B,B,B,S)&=&C_{3 1}^{BS}
=\langle{(\delta{B})^3\delta{S}}\rangle
-3\langle{\delta{B}\delta{S}}\rangle\langle{({\delta{B}})^2}\rangle.
\end{eqnarray}
\end{widetext}

Once we have the definition of mixed-cumulants, the ratio of mixed-cumulants can be calculated.
\begin{figure}[htp]
 \begin{center}
  \includegraphics[scale=0.55]{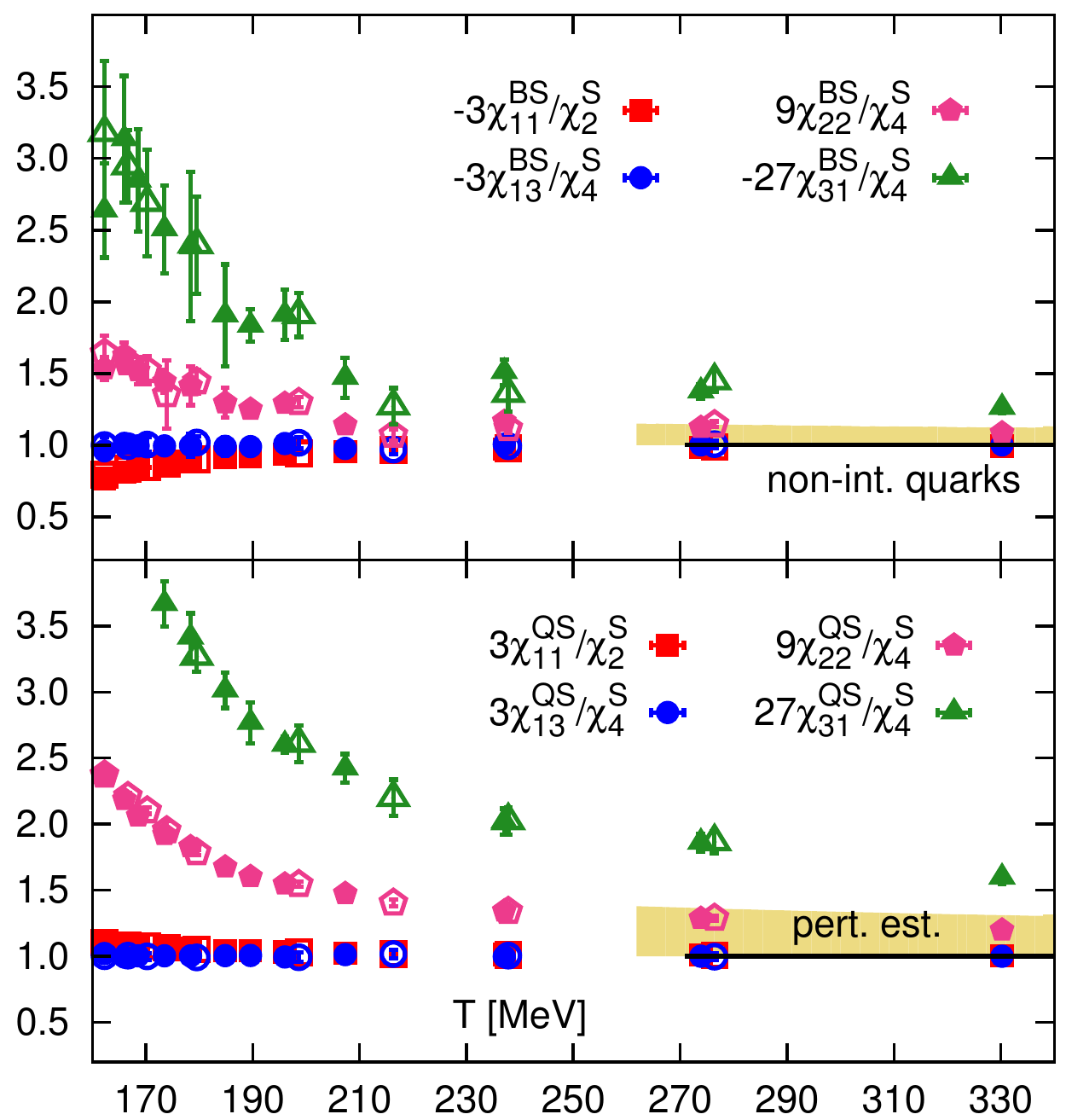}   \caption{Baryon-strangeness (top) and electric charge-strangeness correlation (bottom), properly scaled by the strangeness fluctuations and the power of the fractional baryonic and electric charges [see Eq.~\ref{eq:2}]. The figure is from ~\cite{1304.7220}.}\label{fig:1}
 \end{center}
\end{figure}
A system in thermal equilibrium (for a grand-canonical ensemble) can be characterised by its pressure~\cite{1404.6511}.
The dimensionless pressure of a hadron resonance gas is expressed through the logarithm of the QCD partition function~\cite{PLB738}:
\begin{eqnarray}
\frac{P}{T^4}&=&\frac{1}{VT^{3}}ln[Z(V,T,\mu_{B},\mu_{S},\mu_{Q})]\nonumber\\
&=&\frac{1}{2\pi}\sum_{i\in{X}}g_{i}(\frac{m_{i}}{T})^2K_{2}(\frac{m_{i}}{T})\nonumber\\
&\times &{cosh(B_{i}\hat{\mu}_{B}+Q_{i}\hat{\mu}_{Q}+S_{i}\hat{\mu}_{S})}.
\end{eqnarray}
where $g_{i}$ is the degeneracy factor for hadrons of mass $m_{i}$, and $\hat{\mu}_{q}\equiv\frac{\mu_q}{T}$, with $q=B_{i}$, $S_{i}$, $Q_{i}$ denote the net-baryon number, net-strangeness and the net-charge, and $\mu_{B}$, $\mu_{S}$, $\mu_{Q}$ are the corresponding chemical potentials respectively. For simplicity, we have set the electric charge chemical potential $\hat{\mu}_Q=0.$  These dimensionless fluctuations and correlations of conserved charges (net-baryon number $B$, net-strangeness $S$) are formally equivalent to the Taylor expansion coefficients of the pressure with the respective chemical potentials~\cite{1606.06538,202302,1307.6255}:
\begin{eqnarray}\label{eq:susceptibilites}
\chi_{mn}^{BS}(T,\vec{\mu})=\frac{\partial^{(m+n)}[P(\hat{\mu}_B,\hat{\mu}_S)/T^4]}{\partial\hat{\mu}_B^m\hat{\mu}_S^n}|_{\mu_{B}=\mu_{S}=0}.
\end{eqnarray}
where $\hat{\mu}_B=\frac{\mu_B}{T}$ and $\hat{\mu}_S=\frac{\mu_S}{T}$ are the dimensionless baryon and strangeness chemical potentials, $\chi_{0n}^{BS}\equiv\chi_n^S$ and $\chi_{m0}^{BS}\equiv\chi_m^B$. They are also called generalised susceptibilities. All these derivatives are evaluated at  $\mu_B=\mu_S=0$, the expectation values of all net charge numbers vanish, i.e., $\langle{B}\rangle=\langle{S}\rangle=0$~\cite{PRD034509}.
Theoretically, the mixed-cumulants of these conserved quantities are connected to the corresponding susceptibilities by : 
\begin{eqnarray}
C_{mn}^{BS}=VT^3\chi_{mn}^{BS}(T,\vec{\mu}).
\end{eqnarray}
where the $V,~T$ denotes respectively, the volume and temperature of the system.\\
If above the transition temperature the quarks can be well described by  uncorrelated quasiparticles, then any object which carries strangeness necessarily is a strange quark, then carries baryon number  which in proportion to strangeness, $B_{s}=-\frac{1}{3}S_{s}$\cite{0810.2520}. 
And also, from Eq.~\ref{eq:susceptibilites}, we can derive:
\begin{eqnarray}\label{eq:2}
\frac{\chi_{mn}^{BS}}{\chi_{m+n}^S}=\frac{(-1)^n}{3^m}.
\end{eqnarray} 
where~~$m,n>0$~~and~~$m+n=2,4~.$
Thus one expects baryon number and strangeness to be correlated stronger in a quark-gluon plasma than in a hadron gas. 
The hot medium created in relativistic heavy ion collisions is an expanding system, the spatial volume is changing with time evolution. In order to cancel the effect of the  spatial volume dependence to first order, the ratios $\frac{\chi_{11}^{BS}}{\chi_2^S}$, $\frac{\chi_{13}^{BS}}{\chi_4^S}$, $\frac{\chi_{22}^{BS}}{\chi_4^S}$, $\frac{\chi_{31}^{BS}}{\chi_4^S}$ are constructed.
In order to quantify this and make these ratios to be unity for system where the relevant degree of freedom are quarks, we introduce the following normalised ratios in Eq. \ref{eq:normalized R11}-\ref{eq:normalized R31}:
\begin{widetext}
\begin{eqnarray}
R_{11}^{BS}&=&-3\frac{C_{11}^{BS}}{C_{2}^{S}}=-3\frac{\langle{BS}\rangle-\langle{B}\rangle\langle{S}\rangle}{\langle{S^2}\rangle-\langle{S}\rangle^2},\label{eq:normalized R11}\\
R_{13}^{BS}&=&-3\frac{C_{13}^{BS}}{C_{4}^{S}}
=-3\frac{\langle{BS^3}\rangle -3\langle{BS^2}\rangle\langle{S}\rangle -3\langle{BS}\rangle(\langle{S^2}\rangle -2\langle{S}\rangle^2) +6\langle{B}\rangle\langle{S}\rangle(\langle{S^2}\rangle -\langle{S}\rangle^2) -\langle{B}\rangle\langle{S^3}\rangle}
{\langle{S^4}\rangle -4\langle{S^3}\rangle\langle{S}\rangle -3\langle{S^2}\rangle^2 +12\langle{S^2}\rangle\langle{S}\rangle^2 -6\langle{S}\rangle^4},\label{eq:normalized R13}\\
R_{22}^{BS}&=&9\frac{C_{22}^{BS}}{C_{4}^{S}}\nonumber\\
&=&9\frac{\langle{B^2S^2}\rangle -2\langle{B^2S}\rangle\langle{S}\rangle  -2\langle{BS}\rangle(\langle{BS}\rangle-4\langle{B}\rangle\langle{S}\rangle) -2\langle{BS^2}\rangle\langle{B}\rangle +2\langle{B}\rangle^2(\langle{S^2}\rangle -3\langle{S}\rangle^2) -\langle{B^2}\rangle\langle{S^2}\rangle}
{\langle{S^4}\rangle -4\langle{S^3}\rangle\langle{S}\rangle -3\langle{S^2}\rangle^2 +12\langle{S^2}\rangle\langle{S}\rangle^2 -6\langle{S}\rangle^4}, \label{eq:normalized R22}\\
R_{31}^{BS}&=&-27\frac{C_{31}^{BS}}{C_{4}^{S}}
=-27\frac{\langle{B^3S}\rangle -3\langle{B^2S}\rangle\langle{B}\rangle 
-3\langle{BS}\rangle(\langle{B^2}\rangle -2\langle{B}\rangle^2)
+6\langle{B}\rangle\langle{S}\rangle(\langle{B^2}\rangle -\langle{B}\rangle^2)
-\langle{B^3}\rangle\langle{S}\rangle}
{\langle{S^4}\rangle -4\langle{S^3}\rangle\langle{S}\rangle -3\langle{S^2}\rangle^2 +12\langle{S^2}\rangle\langle{S}\rangle^2 -6\langle{S}\rangle^4}. \label{eq:normalized R31}
\end{eqnarray}
\end{widetext}

At vanishing chemical potentials, the mean values of the conserved charges are vanishing $\langle{B\rangle}=\langle{S\rangle}=0$, the above equations can be simplified as:
\begin{eqnarray}
 R_{11}^{BS}|_{\mu_{B},\mu_{S}=0}&=&-3\frac{\langle{BS}\rangle}{\langle{S^2}\rangle},\\
 R_{13}^{BS}|_{\mu_{B},\mu_{S}=0}&=&-3\frac{\langle{BS^3}\rangle -3\langle{BS}\rangle\langle{S^2}\rangle}{\langle{S^4}\rangle -3\langle{S^2}\rangle^2},\\
 R_{22}^{BS}|_{\mu_{B},\mu_{S}=0}&=&9\frac{\langle{B^2S^2}\rangle -3\langle{B^2}\rangle\langle{S^2}\rangle}{\langle{S^4}\rangle -3\langle{S^2}\rangle^2},\\
 R_{31}^{BS}|_{\mu_{B},\mu_{S}=0}&=&-27\frac{\langle{B^3S}\rangle -3\langle{BS}\rangle\langle{B^2}\rangle}{\langle{S^4}\rangle -3\langle{S^2}\rangle^2}.
\end{eqnarray}
For illustration and discussion purpose, the formula with $\mu_{B}$=$\mu_{S}$=0 are just to demonstrate a special case, which is not used in the calculations presented subsequently. 

\subsection{Baryon-Strangeness Correlations in the Quark Gluon Plasma}
At high temperature, the basic degrees of freedom are weakly interacting quarks and gluons.
The quark operators $u, d, s$ represent the net-quark  number of up, down and strange quarks. We expressed $R_{11}^{BS}$, $R_{13}^{BS}$, $R_{22}^{BS}$ and $R_{31}^{BS}$ in terms of quark degrees of freedom, noting that the baryon number of a quark is $\frac{1}{3}$ and the strangeness of a strange quark is negative one\cite{621(2016)}.
In terms of quark flavours, the various order ratios $R_{11}^{BS}$, $R_{13}^{BS}$, $R_{22}^{BS}$, $R_{31}^{BS}$ can be written as:
\begin{widetext} 
\begin{eqnarray}\label{eq:4}
R_{11}^{BS}|_{\mu_{B},\mu_{S}=0}&=&\frac{\langle{(u+d+s)s}\rangle}{\langle{s^2}\rangle}=1+\frac{\langle{us}\rangle+\langle{ds}\rangle}{\langle{s^2}\rangle},\\
R_{13}^{BS}|_{\mu_{B},\mu_{S}=0}&=&\frac{\langle{(u+d+s)s^3}\rangle -3\langle{(u+d+s)s}\rangle\langle{s^2}\rangle}
{\langle{s^4}\rangle -3\langle{s^2}\rangle^2}
=1+\frac{\langle{us^3}\rangle +\langle{ds^3}\rangle -3(\langle{us}\rangle +\langle{ds}\rangle)\langle{s^2}\rangle}
{\langle{s^4}\rangle -3\langle{s^2}\rangle^2},\\
R_{22}^{BS}|_{\mu_{B},\mu_{S}=0}&=&\frac{\langle{(u+d+s)^2s^2}\rangle -3\langle{(u+d+s)^2}\rangle\langle{s^2}\rangle}{\langle{s^4}\rangle -3\langle{s^2}\rangle^2}\nonumber\\
&=&1+\frac{\langle{(u+d)^2s^2}\rangle +2\langle{(u+d)s^3}\rangle -3[\langle{(u+d)^2}\rangle +2\langle{(u+d)s}\rangle]\langle{s^2}\rangle}
{\langle{s^4}\rangle -3\langle{s^2}\rangle^2},\\
R_{31}^{BS}|_{\mu_{B},\mu_{S}=0}&=&\frac{\langle{(u+d+s)^3s}\rangle -3\langle{(u+d+s)s}\rangle\langle{(u+d+s)^2}\rangle}{\langle{s^4}\rangle -3\langle{s^2}\rangle^2}\nonumber\\
&=&1+\frac{\langle{(u+d)^3s}\rangle +3\langle{(u+d)^2s^2}\rangle -3\langle{(u+d)^2}\rangle\langle{s^2}\rangle +3\langle{(u+d)s^3}\rangle }
{\langle{s^4}\rangle -3\langle{s^2}\rangle^2}\nonumber\\
&-&\frac{-3\langle{(u+d)s}\rangle[\langle{(u+d)^2}\rangle +2\langle{(u+d)s}\rangle +3\langle{s^2}\rangle]}
{\langle{s^4}\rangle -3\langle{s^2}\rangle^2}.
\end{eqnarray}
\end{widetext}

For uncorrelated quark flavours, off-diagonal susceptibilities are vanishing compared to the diagonal susceptibilities\cite{054508(2005), 085016(2002)}, susceptibilities inclusive of strangeness are smaller than those which involve lighter flavours\cite{054901(2006),034501,114014}, then we have
\begin{eqnarray}
\langle{us}\rangle &=&\langle{ds}\rangle=0,\\
\chi_{us}&=&\chi_{ds}\leq\chi_{ud}\ll\chi_{s}\leq\chi_{d}=\chi_{u},
\end{eqnarray}
and thus 
\begin{eqnarray}
R_{mn}^{BS}|_{\mu_{B},\mu_{S}=0}=1.
\end{eqnarray}
where $m, n >0$ and m+n=2, 4.

In the non-interacting massless quark gas all these observables are unity (shown by the lines at high temperatures in Figure~\ref{fig:1}).
 At high temperatures, the shaded regions indicate the ranges of values for these ratios as predicted for the weakly interacting quasi-quarks from the re-summed Hard Thermal Loop (HTL) perturbation theory at the one-loop order~\cite{074003,JHEP05}.

\subsection{Baryon-Strangeness Correlations in Free Hadron Resonance Gas}

It is well known that at low temperature and chemical potentials the QCD can be modelled as a gas of uncorrelated hadrons, where interactions are included through resonances~\cite{QGP3}. These states are taken from Particle Data Book~\cite{PDG}. As the interactions between hadrons are suppressed, the contributions of individual hadrons to thermodynamics can be regarded as free Boltzmann gas~\cite{Asakawa}.
For a gas of uncorrelated hadron resonances, from Eq.~\ref{eq:normalized R11}-\ref{eq:normalized R31}, the numerator receives contributions from only (strange) baryons and anti-baryons, while the denominator receives contributions also from strange mesons~\cite{0505052,QGP3}.
Eq. \ref{eq:normalized R11} can be written as:
\begin{eqnarray}
R_{11}^{BS}\approx3\frac{\langle{\Lambda}\rangle+\langle{\bar{\Lambda}}\rangle+...+3\langle{\Omega^{-}}\rangle+3\langle{\bar{\Omega}^{+}}\rangle}{\langle{K^0}\rangle+\langle{\bar{K}^0}\rangle+...+9\langle{\Omega^{-}}\rangle+9\langle{\bar{\Omega}^{+}}\rangle}.
\end{eqnarray}

Ref \cite{1304.7220} gives the Lattice QCD results for the appropriate combinations of up to fourth order cumulants of net-strangeness fluctuations and their correlations with net-baryon number and electric charge fluctuations and shown in Fig.~\ref{fig:1}.
The ratios of the second order correlations $\chi_{11}^{BS}/\chi_2^S$ and $\chi_{11}^{QS}/\chi_2^S$ are much closer to the expectations for weakly interacting quasi-quarks, differing only around T-1.25$T_{c}$. while the HTL perturbative expansion for ratios involving one derivate of the baryonic/electric charges ($\chi_{11}^{XS}/\chi_2^S$ and $\chi_{13}^{XS}/\chi_4^S$ )~\cite{PLB523} starts differing from the non-interacting quark gas limit~\cite{1304.7220}. It is the same for those involving higher derivatives of the baryonic /electric charges($\chi_{22}^{XS}/\chi_2^4$ and $\chi_{31}^{XS}/\chi_4^S$). 

 \begin{figure*}
 \hspace{0.2cm}
\includegraphics[scale=0.8]{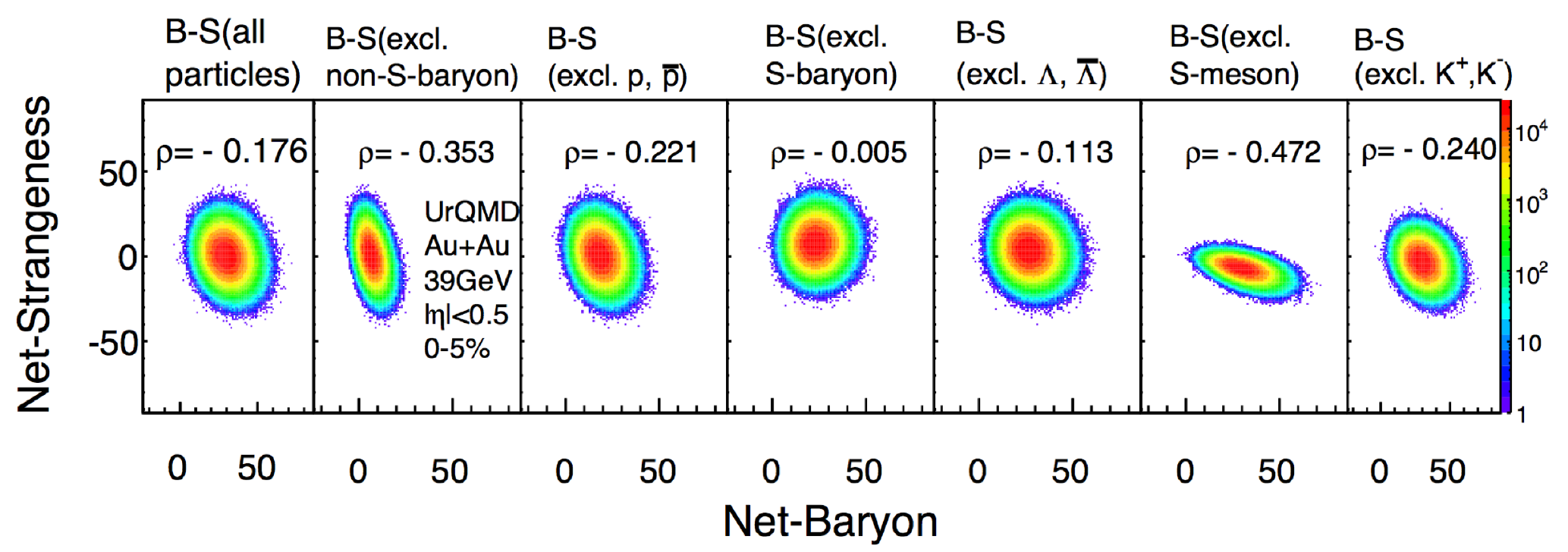}
 \includegraphics[scale=0.85]{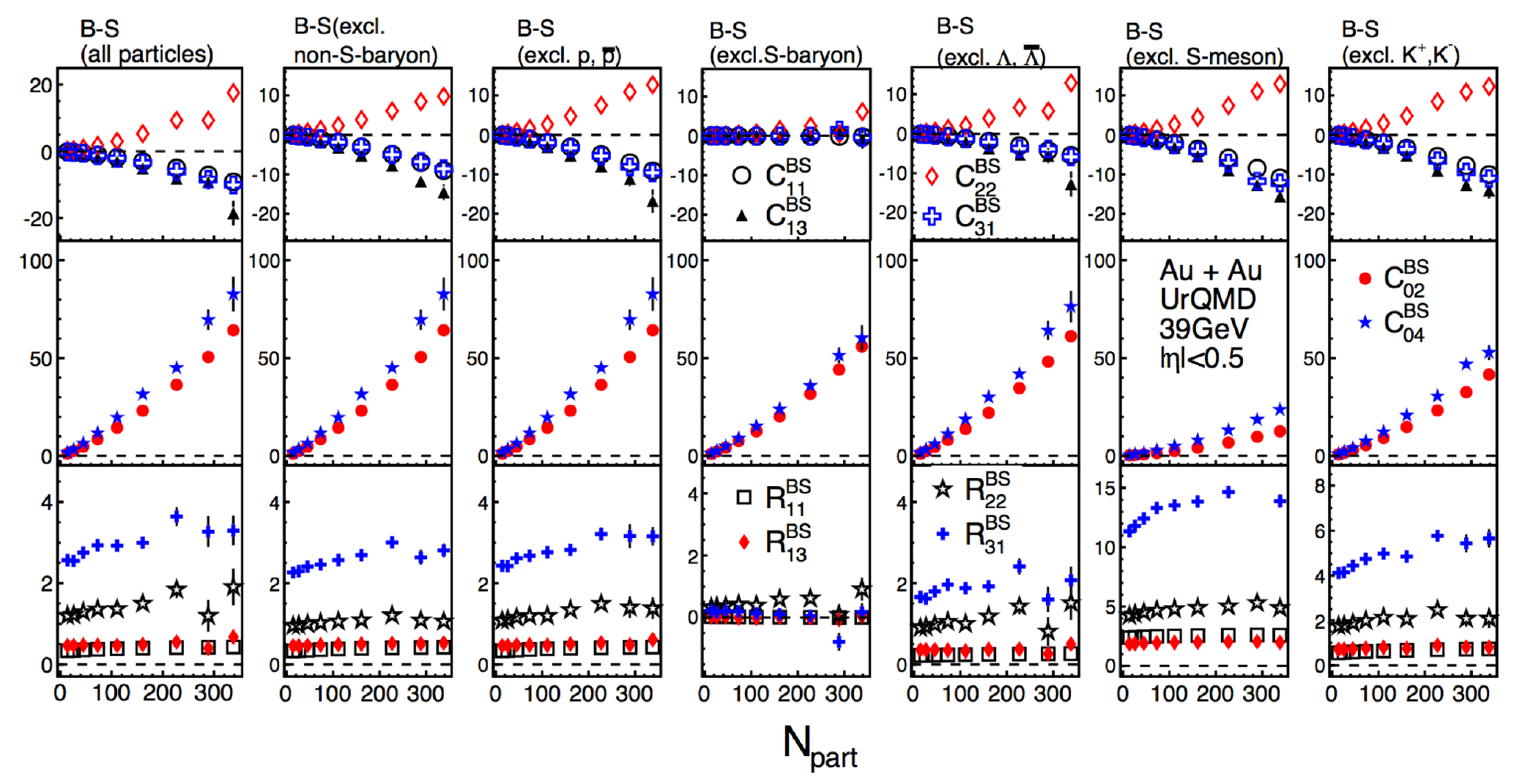}
\caption{Top panel: The baryon-strangeness correlations for the cases \ref{item:4}-\ref{item:10} in the most central (0-5\%) Au+Au collisions at $\sqrt{s_{NN}}$ = 39 GeV from UrQMD model. Bottom panel: The centrality dependence of various order mixed-cumulants ($C_{11}^{BS}$, $C_{13}^{BS}$, $C_{22}^{BS}$, $C_{31}^{BS}$, $C_{02}^{BS}$, $C_{04}^{BS}$) and ratios ($R_{11}^{BS}$, $R_{13}^{BS}$, $R_{22}^{BS}$, $R_{31}^{BS}$) of \ref{item:4}-\ref{item:10} at $\sqrt{s_{NN}}$ = 39GeV for Au+Au collisions from UrQMD model.}
\label{fig:FIG3}
\end{figure*}

\begin{figure*}
 \includegraphics[scale=0.42]{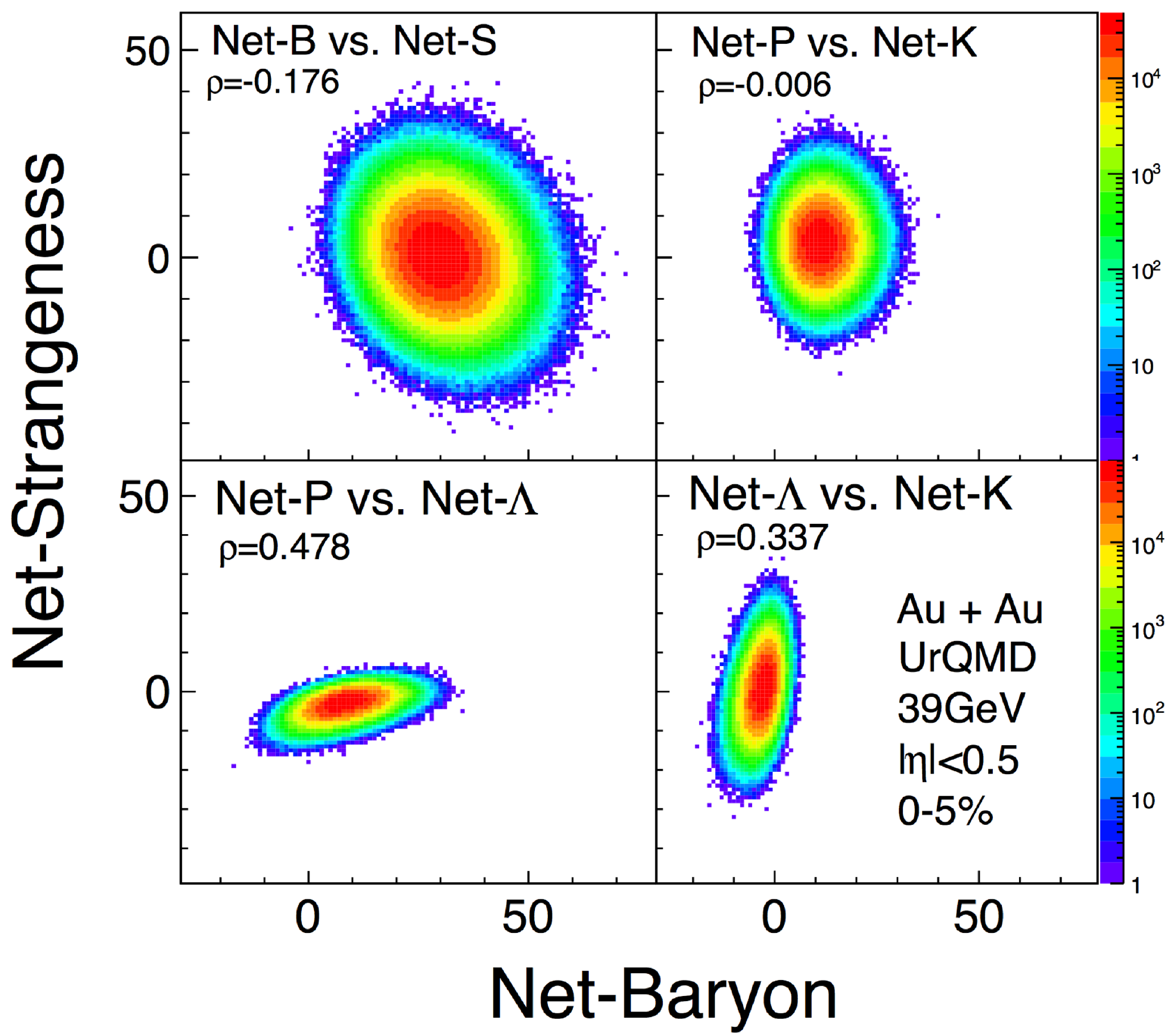}
 \hspace{0.5cm}
\includegraphics[scale=0.4]{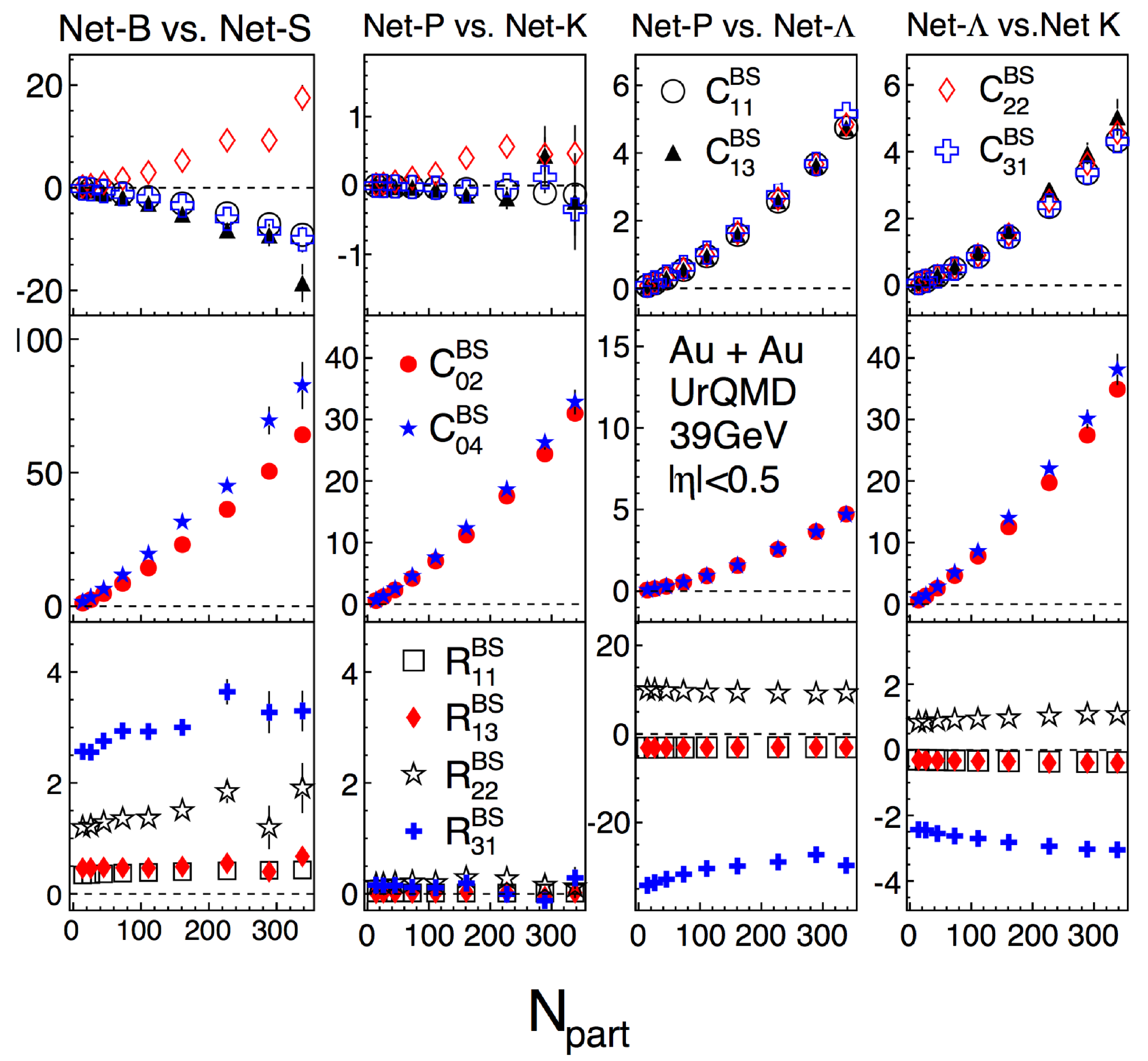}
\caption{Left panel: The correlation between net-baryon and net-strangeness of \ref{item:1}-\ref{item:4} at $\sqrt{s_{NN}}$ = 39GeV for most central (0-5\%) Au+Au collision from UrQMD model. Right panel: The centrality dependence of mixed-cumulants ($C_{11}^{BS}$, $C_{13}^{BS}$, $C_{22}^{BS}$, $C_{31}^{BS}$, $C_{02}^{BS}$, $C_{04}^{BS}$) and ratios ($R_{11}^{BS}$, $R_{13}^{BS}$, $R_{22}^{BS}$, $R_{31}^{BS}$) for \ref{item:1}-\ref{item:4} at $\sqrt{s_{NN}}$ = 39GeV for Au+Au collisions from UrQMD model.}
 \label{fig:FIG2}
\end{figure*}

\section{Results}
In this paper, we performed our calculations with UrQMD model in the version 2.3 for Au+Au collisions at $\sqrt{s_{NN}}$= 7.7, 11.5, 19.6, 27, 39, 62.4, 200 GeV and the corresponding event statistics are 35, 105, 106, 81, 133, 38 and 56 million, respectively~\cite{1606.03900}. The statistical errors are calculated by the formula that are derived from the standard error propagations (see appendix~\ref{error}). To avoid auto-correlations, we define the centralities with the charged particles within the pseudo-rapidity $0.5<|\eta|<1$ and perform the analysis in the pseudo-rapidity range ($|\eta|<0.5 $). For the B-S correlations, we include the particles $p$, $n$, $K^{+}$, $K^{0}$, $\Lambda$, $\Sigma^{-}$,  $\Sigma^{0}$, $\Sigma^{+}$, $\Xi^{-}$, $\Xi^{0}$, $\Omega^{-}$ and their anti-particles. These particles can be classified into strange baryons ($\Lambda$, $\Sigma$, $\Xi$, $\Omega$), non-strange baryons ($p$,$n$) and strange mesons ($K$,$K^{0}$). In order to study the contributions of various hadron species to the baryon-strangeness correlations, we consider the B-S correlations with the following combinations of hadrons: 
\begin{enumerate}[(i)]
\item\label{item:1} Net-P vs. Net-K: Only $p$, $\bar{p}$, $K^{+}$, $K^{-}$ are included;
\item\label{item:2} Net-P vs. Net-$\Lambda$: Only $p$, $\bar{p}$, $\Lambda$, $\bar{\Lambda}$ are included;
\item\label{item:3}Net-$\Lambda$ vs. Net-K: Only $\Lambda$, $\bar{\Lambda}$, $K^{+}$, $K^{-}$ are included; 
\item\label{item:4}B-S (all particles): $p$, $n$, $K^{+}$, $K^{0}$, $\Lambda$, $\Sigma$, $\Xi$, $\Omega$ and their anti-particles are included.
\item\label{item:5}B-S (excl. strange baryon ): Only $p$, $n$, $K^{+}$, $K^{0}$ are included;
\item\label{item:6}B-S (excl. $\Lambda$, $\bar{\Lambda}$): Only $p$, $n$, $K^{+}$, $K^{0}$,  $\Sigma$, $\Xi$, $\Omega$ and their anti-particles are included.
\item\label{item:7}B-S (excl. non-strange baryon): Only $K^{+}$, $K^{0}$, $\Lambda$, $\Sigma$, $\Xi$, $\Omega$ and their anti-particles are included.
\item\label{item:8}B-S (excl. $p$, $\bar{p}$): Only $n$, $K^{+}$, $K^{0}$, $\Lambda$, $\Sigma$,  $\Xi$, $\Omega$ and their anti-particles are included.
\item\label{item:9}B-S (excl. strange meson): Only $p$, $n$, $\Lambda$, $\Sigma$, $\Xi$, $\Omega$ and their anti-particles are included.
\item\label{item:10}B-S (excl. $K^{+}$, $K^{-}$):  Only $p$, $n$, $K^{0}$, $\Lambda$, $\Sigma$, $\Xi$, $\Omega$ and their anti-particles are included.
\end{enumerate}

Figure \ref{fig:FIG3} shows the centrality dependence for the cases \ref{item:4}-\ref{item:10} in the Au+Au collisions at $\sqrt{s_{NN}}$ = 39 GeV from UrQMD model. The top panels of Fig.~\ref{fig:FIG3} displays the two dimension correlations for the cases \ref{item:4}-\ref{item:10} for most central (0-5\%) Au+Au collisions. To quantify the correlation strength, we calculated the standard correlation coefficient for each case. In statistics, the correlation coefficient ($\rho$) is a measure of the linear dependence between two variables $X$ and $Y$, and it is the covariance of the two variables divided by the product of their standard deviations. It is defined as  
\begin{eqnarray}
\rho_{X,Y}=\frac{cov(X,Y)}{\sigma_{X}\sigma_{Y}}=\frac{\langle({X-\langle{X}\rangle})({Y-\langle{Y}\rangle})\rangle}{\sqrt{\langle{X^2}\rangle-\langle{X}\rangle ^2}\sqrt{\langle{Y^2}\rangle-\langle{Y}\rangle ^2}}.\nonumber 
\end{eqnarray}
Where $cov(X,Y)$ is the covariance between random variables $X$ and $Y$. The $\sigma_{X}$ and $\sigma_{Y}$ are the standard deviation of X and Y, respectively. The definition gives a value between +1 and -1, where a value of 1 represents strong correlation, value of 0 indicates no linear correlation, and value of -1  gives perfect anti-correlation. These two dimension plots indicate that most of the cases are anti-correlated and dominated by the strange baryons, such as $\Lambda$. The bottom panels of Fig.~\ref{fig:FIG3} show the centrality dependence of mixed-cumulants ($C_{11}^{BS}$, $C_{13}^{BS}$, $C_{22}^{BS}$, $C_{31}^{BS}$, $C_{02}^{BS}$, $C_{04}^{BS}$) and ratios ($R_{11}^{BS}$, $R_{13}^{BS}$, $R_{22}^{BS}$, $R_{31}^{BS}$) for the cases \ref{item:4}-\ref{item:10} in Au+Au collisions at $\sqrt{s_{NN}}$ = 39 GeV. It is found that the $C_{11}^{BS}$, $C_{13}^{BS}$ and $C_{31}^{BS}$ have negative values, while $C_{22}^{BS}$ have positive values. A strong centrality dependence for the baryon-strangeness (B-S) correlations for the cases \ref{item:6}-\ref{item:10} is observed. By comparing the case \ref{item:4} with \ref{item:5}, we find that the strange baryons have significant contributions to the baryon-strangeness correlations, and the strange mesons ($K^{\pm}$ and $K^{0}$) contribute significantly to strangeness fluctuations of the case~\ref{item:9}~\cite{1606.09573}. When the strange baryons are excluded from the $B$-$S$ correlations, the mixed-cumulant ratios are much smaller than the case with all particles and has values close to zero.  On the other hand, when the strange mesons are excluded from the $B$-$S$ correlations, the values of the ratios ($R_{11}^{BS}$, $R_{13}^{BS}$, $R_{22}^{BS}$, $R_{31}^{BS}$) become larger than the case with all particles.  Furthermore, the results from the case where the non-strange baryons are not included is very close to the results from the case with all particles included. This indicates that the non-strange baryons have little contributions to the baryon strangeness correlations. Since these ratios show weak centrality dependence, we will only consider the most central (0-5\%) collision centralities for the energy dependence.

\begin{figure}
\hspace{-0.2cm}
\includegraphics[scale=0.42]{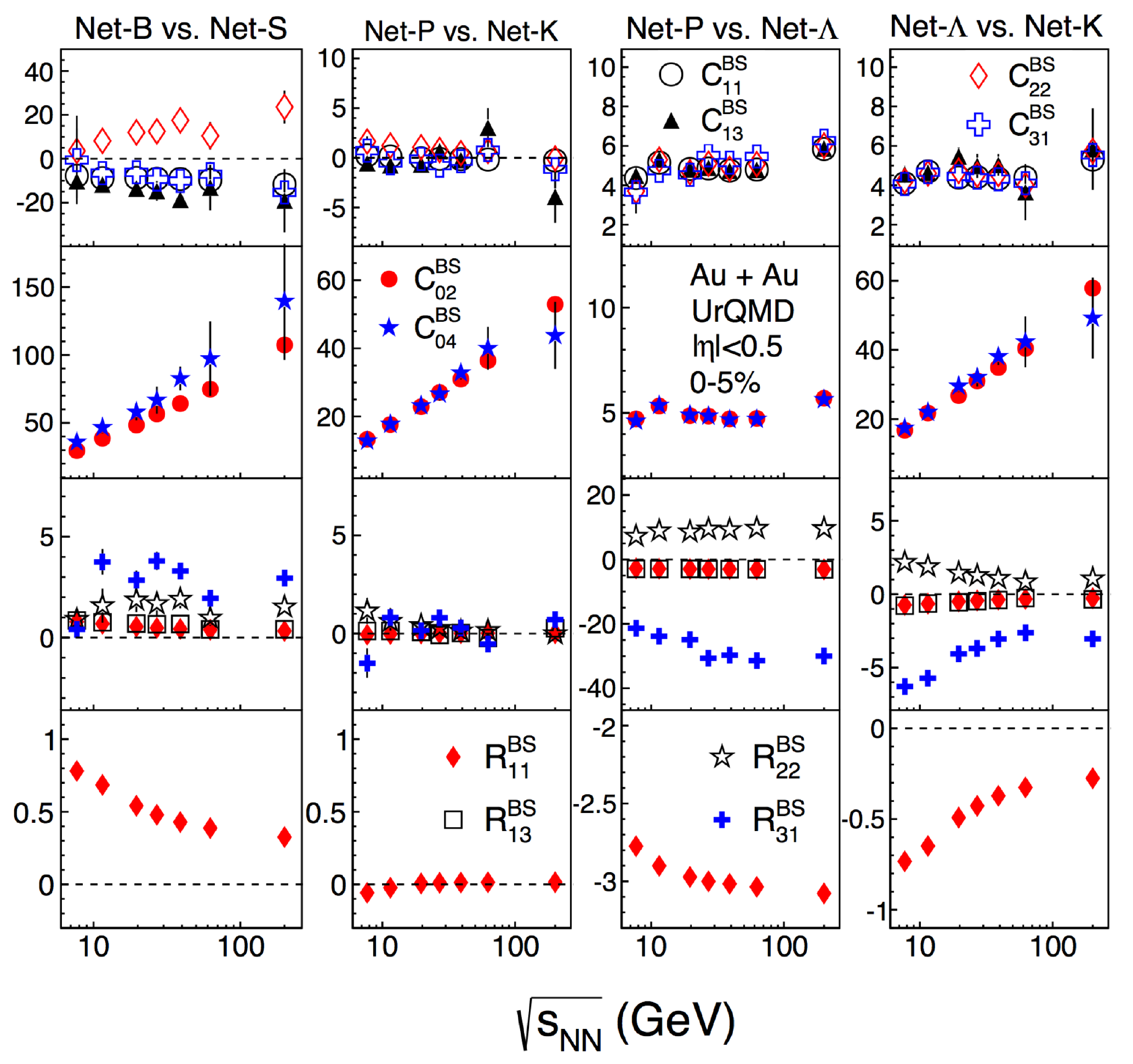} 
\caption{The energy dependence of mixed-cumulants and ratios of the cases \ref{item:1}-\ref{item:4} for most central (0-5\%) Au+Au collisions from UrQMD model.}
\label{fig:FIG4}
\end{figure} 
\begin{figure*}[htpb]
  \begin{center}
    \includegraphics[scale=0.85]{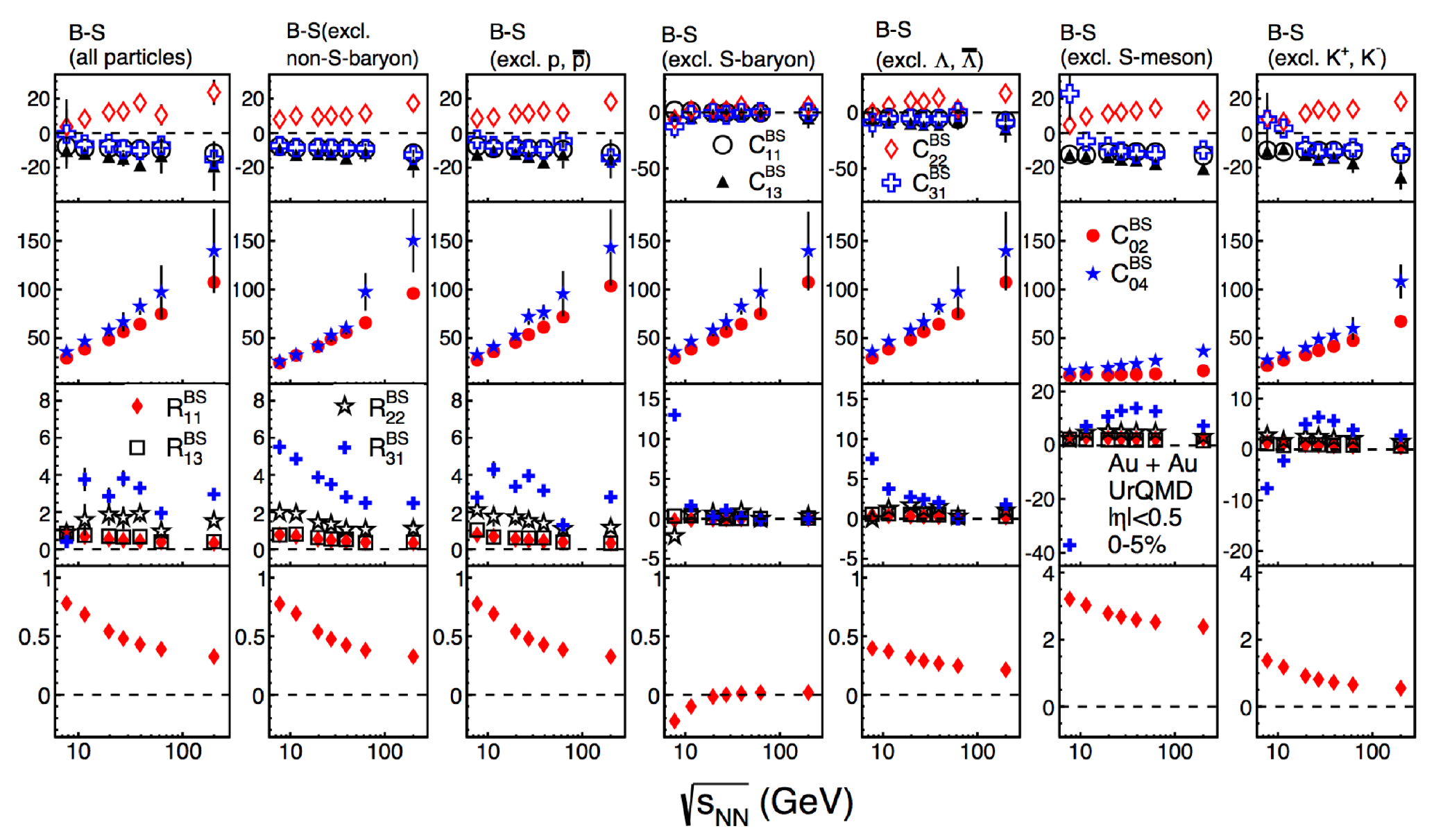}
        \caption{The energy dependence of various order mixed-cumulants ($C_{11}^{BS}$, $C_{13}^{BS}$, $C_{22}^{BS}$, $C_{31}^{BS}$, $C_{02}^{BS}$, $C_{04}^{BS}$) and ratios ($R_{11}^{BS}$, $R_{13}^{BS}$, $R_{22}^{BS}$, $R_{31}^{BS}$)  of \ref{item:4}-\ref{item:10} at $\sqrt{s_{NN}}$ = 7.7, 11.5, 19.6, 27, 39, 62.4 and 200 GeV for the most central (0-5\%) Au+Au collision from UrQMD model. }   
     \label{fig:FIG5}  
   \end{center}
\end{figure*}
Figure \ref{fig:FIG2} shows the centrality dependence of B-S correlations for the cases: \ref{item:1}-\ref{item:4} in Au+Au collisions at $\sqrt{s_{NN}}$ = 39 GeV. The left panels of Fig.~\ref{fig:FIG2} show the correlations for different cases in the most central (0-5\%) Au+Au collisions. The right panels of Fig. \ref{fig:FIG2} show the centrality dependence of mixed-cumulants ($C_{11}^{BS}$, $C_{13}^{BS}$, $C_{22}^{BS}$, $C_{31}^{BS}$, $C_{02}^{BS}$, $C_{04}^{BS}$) and ratios ($R_{11}^{BS}$, $R_{13}^{BS}$, $R_{22}^{BS}$, $R_{31}^{BS}$) for the cases \ref{item:1}-\ref{item:4}.  In the two dimension plots, we find that the net-baryon versus net-strangeness shows strong anti-correlations. If the strange baryons are excluded, for eg. the net-proton and net-kaon correlations are almost independent with each other ($\rho \sim$ 0). It indicates that the strange baryons play an important role and have significant contributions to the $B$-$S$ correlations. In the $B$-$S$ correlations with all particles, the $C_{11}^{BS}$, $C_{13}^{BS}$ and $C_{31}^{BS}$ show negative values and monotonically decrease with increasing of the number of participant nucleons in the collisions, while the $C_{22}^{BS}$ have large and positive values for all of the cases. Interestingly, the ratios $R_{11}^{BS}$, $R_{13}^{BS}$, $R_{22}^{BS}$ and $R_{31}^{BS}$ with all particles included show weak centrality dependence and are comparable with the Lattice QCD results shown in the Fig.~\ref{fig:1}. In the left panels of Fig.~\ref{fig:FIG2}, the net-proton and net-$\Lambda$, net-K and net-$\Lambda$ show finite positive correlations. This can be understood in terms of the baryon stopping and associate productions of the $K^{+}$ and $\Lambda$.  Their mixed-cumulants are positive values, but the ratios $R_{11}^{BS}$, $R_{13}^{BS}$ and $R_{31}^{BS}$ are negative. Furthermore, the variance of the fluctuation of net-$\Lambda$ is much smaller than the variance of net-kaon distributions. This leads to bigger deviation from zero for the proton-$\Lambda$ than kaon-$\Lambda$ correlations.

Figure~\ref{fig:FIG4} shows the energy dependence of the mixed-cumulants and ratios for the cases \ref{item:1}-\ref{item:4} in the most central (0-5\%) Au+Au collisions from UrQMD model. In the $B$-$S$ correlation including all particles, the mixed-cumulants $C_{11}^{BS}$, $C_{13}^{BS}$, $C_{31}^{BS}$ are negative and show weak energy dependence, while the $C_{02}^{BS}$, $C_{04}^{BS}$ and $C_{22}^{BS}$ are positive and monotonically increase with increasing energy. On the other hand, the ratios ($R_{11}^{BS}$, $R_{13}^{BS}$, $R_{22}^{BS}$) show weak energy dependence but the ratio $R_{31}^{BS}$ decreases at low energies. This decreasing trend of the $R_{31}^{BS}$ are also observed in the correlations between net-proton and net-kaon, which might originate from the associate production of $K^{+}$ by the reaction channel $NN \to NYK$, where $N$ is nucleon and $Y$ is hyperon.  The trends of these ratios for the correlations between net-proton and net-$\Lambda$ are opposite to the trends observed in the net-kaon and net-$\Lambda$ correlations. For the two cases, the mixed-cumulant ratios $R_{11}^{BS}$, $R_{13}^{BS}$ and $R_{31}^{BS}$ are negative, $R_{22}^{BS}$ is positive, but higher order ratios $R_{22}^{BS}$ and $R_{31}^{BS}$ have larger value. To demonstrate the energy dependence of the second order ratio $R_{11}^{BS}$ more clearly, we plot $R_{11}^{BS}$ as a function of energy in the bottom panels of Fig.~\ref{fig:FIG4}. It is found that when all particles are considered  or only proton and lambda are included, the $R_{11}^{BS}$ decreases with increasing energy, while we observe different trends for the other two cases.   At low energy, the higher order ratio $R_{31}^{BS}$ becomes small for \ref{item:1} (Net-p vs. Net-K) and \ref{item:4} (Net-B vs. Net-S), the main reason is probably because that at low energy,  the baryon stopping of protons and associate kaon production play important roles. 

Figure~\ref{fig:FIG5} shows the energy dependence of various order mixed-cumulants and ratios of the baryon-strangeness correlations for the cases \ref{item:4}-\ref{item:10} for the most central (0-5\%) Au+Au collisions from UrQMD model. By excluding the non-strange baryons from the $B$-$S$ correlations, we find that the values of ratios $R_{11}^{BS}$, $R_{13}^{BS}$, $R_{22}^{BS}$, $R_{31}^{BS}$ are finite and monotonically increases with decreasing collision energy, it is  comparable with the results from Lattice QCD at low temperature (Lattice QCD results shown in Fig.~\ref{fig:1}). For the $R_{11}^{BS}$, the case including all particles are very close to the one excluding non-strange baryons or protons, while they are different for $R_{31}^{BS}$. It means the non-strange baryons have small effects on the second order ratio $R_{11}^{BS}$, while have large effects on the higher order ratio.  However, if we exclude the strange baryons or $\Lambda$, the values of baryon-strangeness correlations are close to zero at high energies. It indicates that the strange baryons, especially the  $\Lambda$ baryons, carry significant information and play an important role for the $B$-$S$ correlations. Especially, the $R_{31}^{BS}$ show large increase at low energy. It probably because that $\Lambda$ is the lightest strange baryon and contributes most significantly to the baryon-strangeness correlations.  When the $K^{+}$, $K^{-}$ are excluded, there are large impacts on the higher order ratio $R_{31}^{BS}$ at low energies. The ratio becomes negative, which is probably due to the kaon associate production at low energy. There is a strong correlation between protons and kaons as shown in the Fig.~\ref{fig:FIG4}. If we remove the kaons from the $B$-$S$ correlations, the correlations between proton and $\Lambda$ will dominate the $B$-$S$ correlations, which is negative. 

\section{Summary}
We have analysed the centrality and energy dependence of the baryon-strangeness ($B$-$S$) correlations in Au+Au collisions at $\sqrt{s_{NN}}$  = 7.7, 11.5, 19.6, 27, 39, 62.4 and 200 GeV from UrQMD model. The $B$-$S$ correlations are studied via various order mixed-cumulants and corresponding ratios. The B-S correlations are studied via various order mixed-cumulants and corresponding ratios. Such a study provides the following important features: (1) Provides an expectation from a non-CP based model on B-S correlations. The model incorporates standard physics related to baryon number, strangeness number and charge conservation. It also models the baryon stopping at lower beam energies. 
In addition it incorporates resonances, their decay and particle interactions in a hadronic matter. (2) The UrQMD model in this paper is treated as data-like and we have discussed the ways to construct feasible B-S correlations which can be experimentally measured. (3) The B-S correlations results presented in this paper from the analysis of UrQMD model data as one would have carried out in a manner as one would do it using the real experimental data. It has been done using typical acceptance cuts, avoiding auto-correlations while choosing the particle multiplicity used to define the collision centrality and to calculate the correlations, derive the expressions for the statistical errors for such correlations etc. Our results presented
in this paper provides a methodology to look at the experimental data. 	

We have presented the B-S correlations by excluding different hadron species, we have studied ten different cases: ~\ref{item:1}-\ref{item:10}. In these studies, we observed strong correlations between Net-Proton, Net-$\Lambda$ at low energies, which could be related to the baryon stopping and associated production of the hyperons and kaons. The various results of the mixed-cumulants and ratios obtained from UrQMD calculations show a weak centrality dependence.  As far as the energy dependence is concerned, we observe that the strange-baryons and strange-meson have significant contributions to baryon-strangeness correlations and various order mixed-cumulant ratios $R_{13}^{BS}$, $R_{13}^{BS}$, $R_{22}^{BS}$ and $R_{31}^{BS}$, especially at low energy. When strange-baryons are excluded from the $B$-$S$ correlations, the values are relatively small and close to zero. It is found that the model results are also comparable with the Lattice QCD calculations at low temperature. Thus our studies presented in this paper has provided the expectations on the interplay of baryon stopping and kaon associate production to B-S correlations at different collision energies from the UrQMD model. It also provides 
a baselines for the B-S correlations as an observable to search for the signals of the QCD phase transition and critical 
point in heavy-ion collisions.

\section*{Acknowledgements}
The work was supported in part by the MoST of China 973-Project No.2015CB856901, NSFC under grant No. 11575069. BM is supported by DAE, DST and SERB, Govt. of India.

\begin{widetext}

\section{APPENDIX: Statistical Error Calculations} \label{error}

It is well known that the variance of statistic $\Phi(X_{1},X_{1}...X_{m})$ can be expressed in terms of the following~\cite{statistics1}:
\begin{eqnarray}
V(\Phi)=\sum_{i=1,j=1}^{m}(\frac{\partial{\Phi}}{\partial{X_{i}}})(\frac{\partial{\Phi}}{\partial{X_{j}}})Cov(X_{i},X_{j}).
\end{eqnarray}

And the covariance of the multivariate moments can be written as~\cite{statistics2}:
\begin{eqnarray}
Cov(f_{i,j},f_{k,h})&=&\frac{1}{N}(f_{i+k,j+h}-f_{i,j}f_{k,h}).
\end{eqnarray}

where N is the number of events, $f_{i,j}=\langle{B^{i}S^{j}}\rangle$ and $f_{k,h}=\langle{B^{k}S^{h}}\rangle$ are the joint moments of net-baryon and net-strangeness.\\

We use $F_{m,n}$ to present the mixed-central moments,
\begin{eqnarray}
F_{m,n}&=&\langle{(\delta{B})^m(\delta{S})^n}\rangle
=\sum_{i=0}^{m}\sum_{j=0}^{n}C_{m}^{i}C_{n}^{j}(-1)^{m+n-i-j}f_{1,0}^{m-i}f_{0,1}^{n-j}f_{i,j},
\end{eqnarray}

The partial derivation of $F_{m,n}$ to its variable $f_{i,j}$ is:
\begin{eqnarray}
D_{m,n,i,j}&=&\frac{\partial{F_{m,n}}}{\partial{f_{i,j}}} 
=\sum_{i=0}^{m}\sum_{j=0}^{n}C_{m}^{i}C_{n}^{j}(-1)^{m+n-i-j}f_{1,0}^{m-i}f_{0,1}^{n-j},
\end{eqnarray}

The various order mixed-cumulant ratios can be expressed by the terms of joint moments:
\begin{eqnarray}
R_{11}^{BS}&=&-3\frac{C_{11}^{BS}}{C_{02}^{BS}}=-3\frac{F_{1,1}}{F_{0,2}}=-3\frac{f_{1,1}-f_{0,1}f_{1,0}}{f_{0,2}-f_{0,1}^{2}},\\
R_{13}^{BS}&=&-3\frac{C_{13}^{BS}}{C_{04}^{BS}}=-3\frac{F_{1,3}-3F_{1,1}F_{0,2}}{F_{0,4}-3F_{0,2}^{2}}\nonumber\\
&=&-3\frac{f_{1,3}-3f_{1,2}f_{0,1}+6f_{1,1}f_{0,1}^{2}+6f_{1,0}f_{0,2}f_{0,1}-3f_{1,1}f_{0,2}-6f_{1,0}f_{0,1}^{3}-f_{1,0}f_{0,3}}{f_{0,4}-4f_{0,3}f_{0,1}-3f_{0,2}^{2}+12f_{0,2}f_{0,1}^{2}-6f_{0,1}^{4}},\\
R_{22}^{BS}&=&9\frac{C_{22}^{BS}}{C_{04}^{BS}}=9\frac{F_{2,2}-2F_{1,1}^{2}-F_{0,2}F_{2,0}}{F_{0,4}-3F_{0,2}^{2}}\nonumber\\
&=&9\frac{-6f_{0,1}^{2}f_{1,0}^{2}+2f_{0,2}f_{1,0}^{2}+8f_{0,1}f_{1,0}f_{1,1}-2f_{1,1}^{2}-2f_{1,0}f_{1,2}+2f_{2,0}f_{0,1}^{2}-f_{2,0}f_{0,2}-2f_{0,1}f_{2,1}+f_{2,2}}{f_{0,4}-4f_{0,3}f_{0,1}-3f_{0,2}^{2}+12f_{0,2}f_{0,1}^{2}-6f_{0,1}^{4}},\\
R_{31}^{BS}&=&-27\frac{C_{31}^{BS}}{C_{04}^{BS}}=-27\frac{F_{3,1}-3F_{1,1}F_{2,0}}{F_{0,4}-3F_{0,2}^{2}}\nonumber\\
&=&-27\frac{f_{3,1}-3f_{2,1}f_{1,0}+6f_{1,1}f_{1,0}^{2}+6f_{1,0}f_{2,0}f_{0,1}-3f_{1,1}f_{2,0}-6f_{0,1}f_{1,0}^{3}-f_{0,1}f_{3,0}}{f_{0,4}-4f_{0,3}f_{0,1}-3f_{02}^{2}+12f_{0,2}f_{0,1}^{2}-6f_{0,1}^{4}}.
\end{eqnarray}

The partial derivation of $R_{11}^{BS}$ to its variable $f_{i,j}$ is:
\begin{eqnarray}
\frac{\partial{R_{11}^{BS}}}{\partial{f_{ij}}}=\frac{\partial{R_{11}^{BS}}}{\partial{F_{1,1}}}\frac{\partial{F_{1,1}}}{\partial{f_{i,j}}}+\frac{\partial{R_{11}^{BS}}}{\partial{F_{0,2}}}\frac{\partial{F_{0,2}}}{\partial{f_{i,j}}}
=\frac{-3}{C_{02}^{BS}}D_{1,1,i,j}+\frac{3C_{11}^{BS}}{(C_{02}^{BS})^2}D_{0,2,i,j}.
\end{eqnarray}

The variance of observable $R_{11}^{BS}$:\begin{eqnarray}
V(R_{11}^{BS})=\sum_{i,k=0}^{1}\sum_{j,h=0}^{2}\frac{\partial{R_{11}^{BS}}}{\partial{f_{i,j}}}\frac{\partial{R_{11}^{BS}}}{\partial{f_{k,h}}}Cov(f_{i,j},f_{k,h}).
\end{eqnarray}

The error of observable $R_{11}^{BS}$: 
\begin{eqnarray}
Error(R_{11}^{BS})=\sqrt{V(R_{11}^{BS})}.
\end{eqnarray}

The partial derivation of $R_{13}^{BS}$ to its variable $f_{i,j}$ is:
\begin{eqnarray}
\frac{\partial{R_{13}^{BS}}}{\partial{f_{i,j}}}&=&\frac{\partial{R_{13}^{BS}}}{\partial{F_{1,3}}}\frac{\partial{F_{1,3}}}{\partial{f_{i,j}}}
+\frac{\partial{R_{11}^{BS}}}{\partial{F_{1,1}}}\frac{\partial{F_{1,1}}}{\partial{f_{i,j}}}
+\frac{\partial{R_{11}^{BS}}}{\partial{F_{0,4}}}\frac{\partial{F_{0,4}}}{\partial{f_{i,j}}}
+\frac{\partial{R_{11}^{BS}}}{\partial{F_{0,2}}}\frac{\partial{F_{0,2}}}{\partial{f_{i,j}}}\nonumber\\
&=&\frac{-3}{C_{04}^{BS}}D_{1,3,i,j}+\frac{3C_{02}^{BS}}{C_{04}^{BS}}D_{1,1,i,j}
+\frac{3C_{13}^{BS}}{(C_{04}^{BS})^2}D_{0,4,i,j}
+\frac{9C_{11}^{BS}C_{04}^{BS}+18C_{13}^{BS}C_{02}^{BS}}{(C_{04}^{BS})^2}D_{0,2,i,j}.
\end{eqnarray}

The variance of observable $R_{13}^{BS}$:
\begin{eqnarray}
V(R_{13}^{BS})=\sum_{i,k=0}^{1}\sum_{j,h=0}^{4}\frac{\partial{R_{13}^{BS}}}{\partial{f_{i,j}}}\frac{\partial{R_{13}^{BS}}}{\partial{f_{k,h}}}Cov(f_{i,j},f_{k,h}).
\end{eqnarray}

The error of observable $R_{13}^{BS}$:
\begin{eqnarray}
Error(R_{13}^{BS})=\sqrt{V(R_{13}^{BS})}.
\end{eqnarray}

The partial derivation of $R_{22}^{BS}$ to its variable $f_{i,j}$ is:
\begin{eqnarray}
\frac{\partial{R_{22}^{BS}}}{\partial{f_{i,j}}}&=&\frac{\partial{R_{22}^{BS}}}{\partial{F_{2,2}}}\frac{\partial{F_{2,2}}}{\partial{f_{i,j}}}
+\frac{\partial{R_{22}^{BS}}}{\partial{F_{2,0}}}\frac{\partial{F_{2,0}}}{\partial{f_{i,j}}}
+\frac{\partial{R_{22}^{BS}}}{\partial{F_{1,1}}}\frac{\partial{F_{1,1}}}{\partial{f_{i,j}}}
+\frac{\partial{R_{22}^{BS}}}{\partial{F_{0,4}}}\frac{\partial{F_{0,4}}}{\partial{f_{i,j}}}
+\frac{\partial{R_{22}^{BS}}}{\partial{F_{0,2}}}\frac{\partial{F_{0,2}}}{\partial{f_{i,j}}}\nonumber\\
&=&\frac{9}{C_{04}^{BS}}D_{2,2,i,j}+\frac{-9C_{02}^{BS}}{C_{04}^{BS}}D_{2,0,i,j}+\frac{-36C_{11}^{BS}}{C_{04}^{BS}}D_{1,1,i,j}
+\frac{-9C_{22}^{BS}}{(C_{04}^{BS})^{2}}D_{0,4,i,j}
+\frac{9C_{20}^{BS}C_{04}^{BS}+54C_{02}^{BS}C_{22}^{BS}}{(C_{04}^{BS})^{2}}D_{0,2,i,j}.\nonumber\\
\end{eqnarray}

The variance of observable $R_{22}^{BS}$:
\begin{eqnarray}
V(R_{22}^{BS})=\sum_{i,k=0}^{2}\sum_{j,h=0}^{4}\frac{\partial{R_{22}^{BS}}}{\partial{f_{i,j}}}\frac{\partial{R_{22}^{BS}}}{\partial{f_{k,h}}}Cov(f_{i,j},f_{k,h}).\nonumber\\
\end{eqnarray}

The error of observable $R_{22}^{BS}$:
\begin{eqnarray}
error(R_{22}^{BS})=\sqrt{V(R_{22}^{BS})}.
\end{eqnarray}

The partial derivation of $R_{31}^{BS}$ to its variable $f_{i,j}$ is:
\begin{eqnarray}
\frac{\partial{R_{31}^{BS}}}{\partial{f_{i,j}}}&=&\frac{\partial{R_{31}^{BS}}}{\partial{F_{3,1}}}\frac{\partial{F_{3,1}}}{\partial{f_{i,j}}}
+\frac{\partial{R_{31}^{BS}}}{\partial{F_{2,0}}}\frac{\partial{F_{2,0}}}{\partial{f_{i,j}}}
+\frac{\partial{R_{31}^{BS}}}{\partial{F_{1,1}}}\frac{\partial{F_{1,1}}}{\partial{f_{i,j}}}
+\frac{\partial{R_{31}^{BS}}}{\partial{F_{0,4}}}\frac{\partial{F_{0,4}}}{\partial{f_{i,j}}}
+\frac{\partial{R_{31}^{BS}}}{\partial{F_{0,2}}}\frac{\partial{F_{0,2}}}{\partial{f_{i,j}}}\nonumber\\
&=&\frac{-27}{C_{04}^{BS}}D_{3,1,i,j}+\frac{81C_{11}^{BS}}{C_{04}^{BS}}D_{2,0,i,j}
+\frac{81C_{20}^{BS}}{(C_{04}^{BS})^2}D_{1,1,i,j}
+\frac{27C_{31}^{BS}}{(C_{04}^{BS})^2}D_{0,4,i,j}
+\frac{-162C_{02}^{BS}C_{31}^{BS}}{(C_{04}^{BS})^2}D_{0,2,i,j}.\nonumber\\
\end{eqnarray}

The variance of observable $R_{31}^{BS}$:
\begin{eqnarray}
V(R_{31}^{BS})=\sum_{i,k=0}^{3}\sum_{j,h=0}^{4}\frac{\partial{R_{31}^{BS}}}{\partial{f_{i,j}}}\frac{\partial{R_{31}^{BS}}}{\partial{f_{k,h}}}Cov(f_{i,j},f_{k,h}).\nonumber\\
\end{eqnarray}

The error of observable $R_{31}^{BS}$:
\begin{eqnarray}
Error(R_{31}^{BS})=\sqrt{V(R_{31}^{BS})}.
\end{eqnarray}
\end{widetext}

\end{document}